\documentclass[
superscriptaddress,
bibnotes,
amsmath,amssymb,
aps,
prl, 
floatfix,
reprint,
longbibliography
]{revtex4-2}

\usepackage{lineno}
\usepackage{graphicx}

\usepackage{mathrsfs,dsfont,mathtools,bm, braket, amsmath, amsfonts}
\usepackage[dvipsnames]{xcolor}
\usepackage[
    colorlinks,
    linkcolor={blue!80!black},
    citecolor={blue!80!black},
    urlcolor={blue!80!black}
]{hyperref}

\usepackage[babel,kerning=true,spacing=true]{microtype}
\usepackage{verbatim}
\usepackage{soul}

\newcommand{\be}{\begin{equation}}
\newcommand{\ee}{\end{equation}}
\newcommand{\bea}{\begin{eqnarray}}
\newcommand{\eea}{\end{eqnarray}}

\newcommand{\ra}{\rangle}

\newcommand{\lp}{\left(}
\newcommand{\rp}{\right)}

\usepackage{xcolor}

\usepackage[normalem]{ulem} 

\begin{document}

\title{
Miniband Generation by Surface Acoustic Waves
}
\author{Eli Meril}
\author{Unmesh Ghorai}
\author{Tobias Holder}
\author{Rafi Bistritzer}
\email{rafib@tauex.tau.ac.il}
\affiliation{School of Physics and Astronomy, Tel Aviv University, Tel Aviv 69978, Israel}
\date{\today}

\begin{abstract}

We introduce a new class of tunable periodic structures 
formed by launching two obliquely propagating surface acoustic waves on a piezoelectric substrate that supports a two-dimensional material. The resulting \emph{acoustoelectric superlattice} exhibits two salient features. First, its periodicity is widely tunable, spanning a length scale intermediate between moiré superlattices and optical lattices, enabling the formation of narrow, topologically nontrivial energy bands. Second, unlike moiré systems, where the superlattice amplitude is set by intrinsic interlayer tunneling and lattice relaxation, the amplitude of the acoustoelectric potential is externally tunable via the surface acoustic wave power. Using massive monolayer graphene as an example, we demonstrate that varying the frequencies and power of the surface acoustic waves enables \emph{in-situ} control over the band structure of the 2D material, generating flat bands and nontrivial valley Chern numbers, featuring a highly localized Berry curvature. 

\end{abstract}
\maketitle

\emph{Introduction---}
Controlling the periodicity of a material is a powerful means of shaping its physical behavior. 
Prominent examples include photonic ~\cite{photonicCystalsReview2019} and phononic crystals ~\cite{phononicCrystalsReview2010, sonicMaterials2000}, 
optical lattices~\cite{coldAtomLatticesReview2007}, and metamaterials~\cite{3Dmetamaterials2019}. More recently, moiré superlattices realizing periodicities on the order of 10~nm have emerged as a particularly fruitful platform for studying strong correlations and topology in flat bands ~\cite{moireBands2011,moireButterflies_2011,MATBG_Cao_Mott_2018,MATBG_Cao_SC_2018,MATBG_Sharpe_FM_2019, 
MATBG_DiracRivavls_Zondiner_2020,twistedCrI3_Song_2021,MATBG_Pomeranchuk_Rozen_2021,TBG_Review_Allan_Eva_2020,TBG_PhysToday_2024,MATBG_CI_Xie_2019,MATBG_CI_Nuckolls_2020,MATBG_CI_Stepanov_2021,MATBG_CI_Park_2023,MATBG_Felectrons_Song_2022,MATBG_CI_Wang_2024,MATBG_SC_Yankowitz_2019,MATG_SC_Stepanov_2020,MATG_SC_Hao_2021}.

\begin{figure*}[t]
    \centering
    \includegraphics[width=\textwidth]{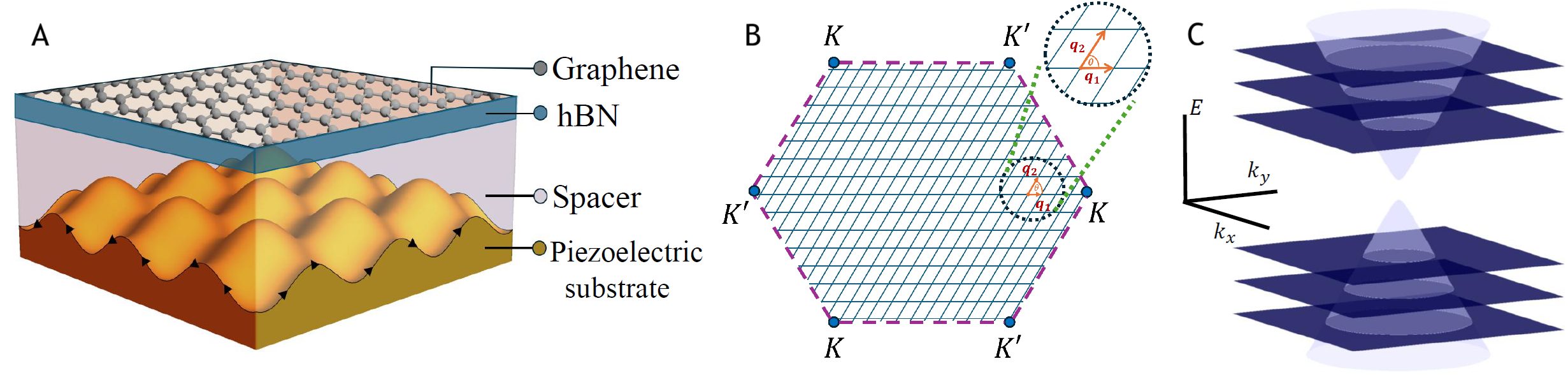}
    \caption{ 
  (A) Schematic of the device. A 2D material is placed on a piezoelectric substrate, separated by a spacer layer. 
  Two SAWs propagating through the substrate, generate a superlattice potential in the 2D material. 
  (B) The SAW wavevectors $\bm{q}_1$ and $\bm{q}_2$ define a mBZ which tiles a K-lattice within the original BZ. (C) The valence and conduction bands of the 2D material split into a set of minibands defined over the mBZ.}
    \label{fig: ASL}
\end{figure*}

In this work, we propose and theoretically investigate a new class of periodic structures: the acoustoelectric superlattice (ASL). As illustrated in Fig.~\ref{fig: ASL}A, an ASL is formed by placing a 2D material on a piezoelectric substrate through which two obliquely propagating surface acoustic waves (SAWs) are launched \cite{SAW_grapheneMinguez_2018, AE_graphene_Mou_2025}. The interference of these SAWs generates a tunable superlattice potential within the 2D material.

\begin{figure}[htp]
    \includegraphics[width=1\columnwidth, height=6.5cm]{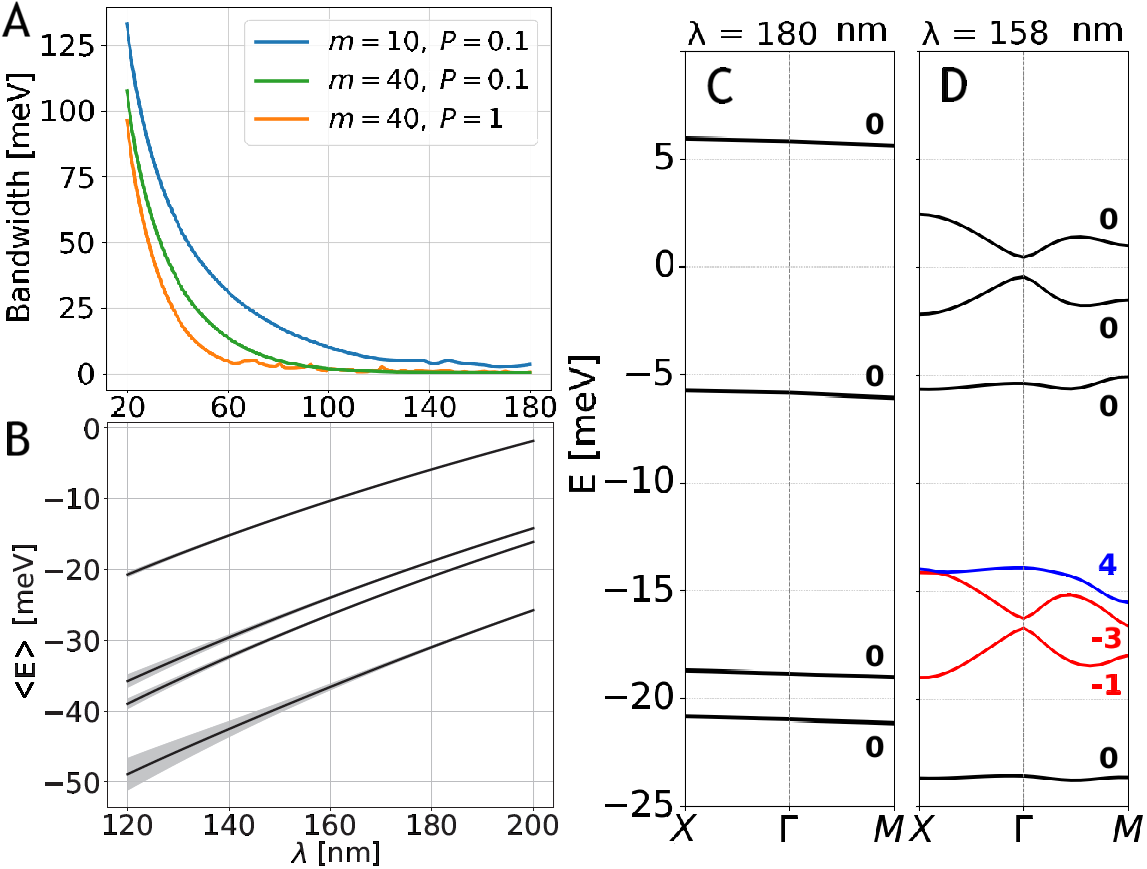}
    \caption{(A) Bandwidth of the highest valence miniband as a function of $\lambda$ for selected $(m,P)$ values  (B) Average energy of top valence minibands as a function of $\lambda$ for $P=0.1$~W/m and $m=40$~meV; the curve width reflects the bandwidth. (C) Atomic-like spectra at $\lambda = 180$~nm, $m=40$~meV, and $P = 0.1 $~W/m. (D) Dispersive minibands with non-zero $C_v$ at $\lambda=158$~nm, $m=20$~meV and $P=0.2$~W/m.}
    \label{fig: dense bands}
\end{figure}

The ASL exhibits two key features.
First, it introduces a new tunable length scale, ranging from tens of nanometers, set by the highest accessible SAW frequencies in the GHz range \cite{SAW_45nm_Nelson_2012PRB, SAW_44G_2023, SAW_resonator_cornell, SAW_LiNbO3_GHz_Japan}, up to the electron mean free path, beyond which the potential is no longer coherently sampled. This range bridges the gap between moiré and optical lattice periodicities and enables the formation of narrow, closely spaced energy bands (see Fig.~\ref{fig: ASL}C and Fig.~\ref{fig: dense bands}).
Second, the frequency, amplitude, phase, and propagation direction of each SAW can be independently controlled. 
This enables the lattice geometry to be tuned independently of the potential amplitude, a capability unavailable in moiré systems, where for a given twist angle, the periodic potential is fixed by interlayer interactions.

Furthermore, ASLs require no rotational alignment, enabling use with 2D materials unsuitable for tear-and-stack~\cite{tearAndStack_Tutuc_2016,tearAndStack_Cao_2016}. They also avoid twist-angle disorder~\cite{Uri2020} and interlayer corrugation which can induce uncontrolled modifications to the electronic structure ~\cite{corrugation_MATBG_2017_Nam,corrugation_MATBG_2019_Carr,corrugation_MATBG_2019_Lucignano}.

In the following, we investigate the electronic structure of massive graphene subjected to an acoustoelectric superlattice, focusing on a single valley. The framework, however, is general and applies broadly to 2D materials.

\emph{Model---} The SAW in the piezoelectric substrate couples to the 2D system via two distinct mechanisms. The first is the piezoelectric effect, where the electric field generated by the SAWs directly interacts with the 2D electrons. The second is the deformation potential, arising from strain transferred from the vibrating substrate to the 2D layer. 

In the following, we focus exclusively on the piezoelectric effect.
The deformation potential can be selectively suppressed by mechanically decoupling the 2D layer from the substrate, either by suspending it or by inserting a stiff spacer that blocks strain transmission (see Fig.~\ref{fig: ASL}A). Importantly, the piezoelectric field remains effective, as its evanescent decay into the spacer is negligible provided the spacer is thinner than the SAW wavelength and given that its dielectric constant is sufficiently small.
The resulting electric potential 
$ V(\bm{r},t) = g_1 \cos\big( \bm{q}_1 \cdot \bm{r} - \Omega_1 t \big)
+ g_2 \cos\big( \bm{q}_2 \cdot \bm{r} - \Omega_2 t \big) $
oscillates in both space and time. It describes a moving superlattice characterized by the same wavevectors $\bm{q}_j$ and frequencies $\Omega_j= v_s q_j$ as the underlying SAWs, where $v_s$ is the substrate sound velocity for Rayleigh waves. The amplitude of each potential component is approximated by (see Supplementary Information~\footnote{In the Supplementary Information further details can be found about the transformation to the moving frame, the SAW-induced potential, and the approximate ten-band model.}),
\be
g_j = \alpha_{\mathrm{PE}} \sqrt{P_j \lambda_j}, 
\qquad 
\alpha_{\mathrm{PE}} = \sqrt{ \frac{K_{\mathrm{R}}^2 e^2}{\pi v_s \varepsilon_{\mathrm{eff}}} }
\label{gPE}
\ee
where $P_j$ is the SAW power per unit width, $\lambda_j = 2\pi/q_j$ is the wavelength, and $e$ is the electron charge. The coupling coefficient $\alpha_{\mathrm{PE}}$ encodes the substrate parameters, including the electromechanical coupling coefficient
$K_{\mathrm{R}}^2$ and the effective surface dielectric constant $\varepsilon_{\mathrm{eff}}$ ~\cite{SAW_Zhang}. 

As in moiré systems, it is useful to analyze the Hamiltonian in momentum space using the extended-zone scheme \cite{moireBands2011}. In this approach the full momentum space is partitioned into equivalence classes, or K-lattices, indexed by a crystal momentum 
$\bm{k}$ in the mini Brillouin zone (mBZ). Each K-lattice consists of momenta $ \bm{k+G} $ where $\{  \bm{G} \} = \{ i \bm{q}_1 +j \bm{q}_2 \}$ are the reciprocal lattice vectors of the superlattice (see Fig.~\ref{fig: ASL}B). The periodic potential couples only states within the same K-lattice. The effective Hamiltonian for a given K-lattice takes the form
\begin{align}
H_{\bm{k}}(t) 
  &= \sum_{\bm G } \Bigg\{ 
        h_{k+G}\, c_{\bm k+G}^\dagger c_{\bm k+G} 
        + \Big[ 
             \frac{g_1}{2} e^{-i\Omega_1 t} c_{\bm k+G+q_1}^\dagger c_{\bm k+G}  \nonumber \\
  &\qquad\qquad\quad 
        +  \frac{g_2}{2} e^{-i\Omega_2 t} c_{\bm k+G+q_2}^\dagger c_{\bm k+G}
        + \mathrm{h.c.} 
        \Big] 
     \Bigg\} 
\label{Hk}
\end{align}
where $c_{\bm k}$ annihilates an electron of momentum $\bm{k}$ from the K-lattice, h.c. denotes the Hermitian conjugate, and internal degrees of freedom (e.g., spin, sublattice) are suppressed for brevity.

We focus on low-energy phenomena in massive graphene and approximate $h_{\bm{k}}$ by the continuum model $ h_{\bm{k}} = v(\tau \sigma_x k_x + \sigma_y k_y) + m \sigma_z$,
where $\bm{k}$ denotes momentum measured from the center of the $K$ or $K'$ valleys, $v$ is the Fermi velocity, $\sigma_i$ are the three Pauli matrices acting on the sublattice degree of freedom, and $\tau = \pm 1$ labels the valley index (hereafter $\hbar=1$). The mass term $m\sigma_z$ breaks the chiral (sublattice) symmetry that protects the gapless Dirac point, thereby opening a band gap. In graphene, such a mass typically originates from alignment with a hexagonal boron nitride substrate; its magnitude depends sensitively on the local stacking configuration and can be as large as a few tens of meV ~\cite{diracMass_Hunt_2013, diracMass_Woods_2014, diracMass_Jung_2015,diracMass_Liu_2022}. 
The massive Dirac Hamiltonian carries a valley-dependent Chern number of $\pm 1/2$ in each band, reflecting its nontrivial topology. 
As in moiré superlattices, the atomic-scale lattice becomes irrelevant, and the effective periodicity is dictated entirely by the superlattice potential.
\begin{figure*}[t!]
    \centering
    \includegraphics[width=\textwidth]{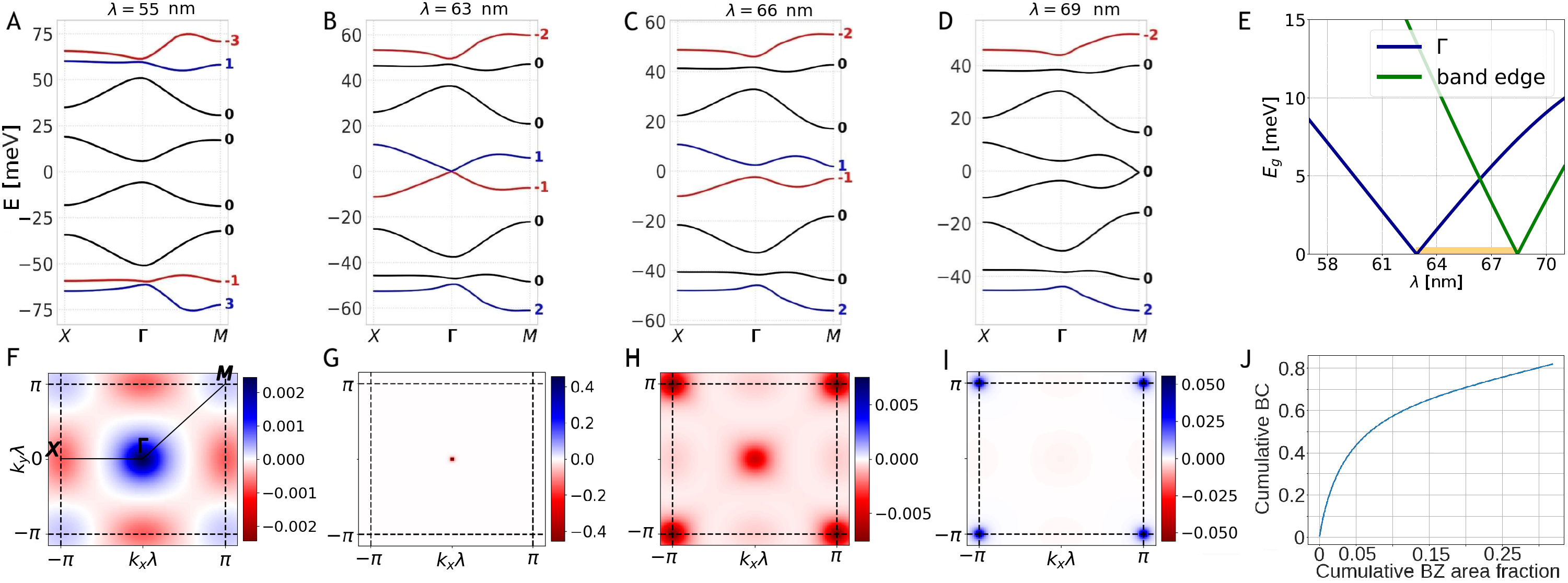}    \caption{
    Evolution of the minibands (A–D) and corresponding Berry curvature of the top valence miniband (F–I) at the K valley as the SAW wavelength is varied for $m = 20$~meV and $P = 1$~W/m. Minibands are labeled by their valley Chern numbers. At small wavelengths, the conduction and valence minibands near charge neutrality ($E = 0$) have zero Chern number. As $\lambda$ increases, a band inversion first occurs at the Dirac point $\Gamma$, followed by a second inversion at the $M$ point. (E) Direct gaps between top valence and bottom conduction minibands at $\Gamma$ and the band edge show a maximum value of $\sim 5$~meV at $\lambda \approx 66$~nm. (J) Even at this wavelength, where the gap is maximal, the Berry curvature (BC) remains highly localized in momentum space.}
    \label{fig: band evolution}
\end{figure*}

\emph{Moving frame---}  The Hamiltonian $H_{\bm{k}}(t)$ can be rendered time-independent via the unitary transformation $U_{\bm{k}}(t) = \exp{\lp i\Delta_{\bm{k}} t \rp}$ with
\begin{equation}
\Delta_{\bm{k}} = \sum_{\bm G} \omega_{\bm G} 
c_{\bm k+G}^\dagger c_{\bm k+G},
\label{Delta_k}
\end{equation}
where $\omega_{\bm G} = i \Omega_1 + j \Omega_2$ with $i,j$ the integers that relate $\bm{G}$ with $\bm{q}_1$ and $\bm{q}_2$.
This transformation remains valid even when the drive frequencies $\Omega_1$ and $\Omega_2$ are incommensurate.
The wavefunction transforms accordingly as $|\psi'_{\bm{k}}(t)\ra = U_{\bm{k}}(t)|\psi_{\bm{k}}(t)\ra$ (see ~\cite{Note1}).
The transformed Hamiltonian is
\begin{equation}
H'_{\bm{k}} = H_{\bm{k}}(0) - \Delta_{\bm{k}}.
\label{Hk_movingFrame}
\end{equation}
Its eigenvalues define a set of quasienergy bands used to analyze Berry curvature and topological response. 
When $\Omega_1$ and $\Omega_2$  are commensurate, $H_{\bm{k}}(t)$ is periodic in time, and the transformation reduces to the standard Floquet representation~\cite{Poertner2020,Wang2023a};
for incommensurate drives $H'_{\bm{k}}(t)$ provides an effective time-independent description whose spectrum plays the role of generalized quasienergy bands.

The Hamiltonian (\ref{Hk_movingFrame}) in the moving frame consists of two distinct contributions. The first is a static component corresponding to the potential as seen in the moving frame. This term mimics a spatial modulation akin to that produced by a lithographically patterned substrate \cite{periodicStrain_Forsythe_2018, periodicStrain_Ruiz_2022}. The second term, $\Delta_{\bm{k}}$, varies linearly across the K-lattice and acts analogously to a uniform electric field in momentum space. It breaks time-reversal (TR), chiral, and inversion symmetries. 

Within each valley, $\Delta_{\bm{k}}$ yields a parametrically small energy shift of order of $\Omega$. This follows from the effective truncation of the sum in Eq. (3) at order $g/vq$, and the oddness of $\omega_G$ with respect to the origin of the K-lattice. 

Although $\Delta_{\bm{k}}$ is linear across the K-lattice and might appear to lift the valley degeneracy, its effect is a gauge-dependent quasienergy offset that cancels in observables. A naive reading of $H'_{\bm{k}}$ suggests a relative valley shift of order $2v_s k_D \approx 40$ meV. However, 
the large quasienergy shift is exactly undone by $U_{\bm{k}}^\dagger(t)$  upon transforming the wavefunction back to the lab frame. Put differently, shifting $\Delta_{\bm{k}}$ by 
$\Delta_0 = \epsilon_0 \sum_{\bm{G}}  c_{\bm{k+G}}^\dagger c_{\bm{k+G}}$ leaves $|\psi_{\bm{k}}(t) \ra$ invariant, mirroring Floquet quasienergy folding.

Consequently, while SAWs weakly break TR symmetry, the valley degeneracy is preserved to an excellent approximation in the absence of additional perturbations. This degeneracy can, however, be deliberately lifted by introducing explicit TR-breaking mechanisms, for example via proximity to a magnetic substrate~\cite{ valleyDeg_Norden_2019,valleyDeg_Zhao_2017}.

\emph{Band reconstruction---} The seemingly simple heterostructure in Fig.~\ref{fig: ASL}A supports a remarkably rich variety of band structures, governed by a few tunable parameters: Dirac mass, SAW wavelengths, and SAW power. Variation of these controls the mBZ size, miniband widths, and the topological characteristics of the minibands.  

To gain insight into how ASLs reshape the band structure of 2D materials, we examine a symmetric configuration of two identical SAWs propagating along orthogonal directions, such that $q_1=q_2 \equiv q$ and $g_1=g_2 \equiv g$.

For numerical estimates, we consider a hBN spacer and a LiNbO$_3$ substrate, a natural choice due to its strong piezoelectric response 
~\cite{SAW_LiNbO3_GHz_Japan}. We adopt representative parameters of $v_s=3488$~m/s, $\varepsilon_{\mathrm{eff}}=14 \varepsilon_0$, $\varepsilon_0$ being the vacuum permittivity, and $K_{\mathrm{R}}^2=0.05$. 
The highest SAW power considered is 1 W/m. For a device of lateral size $\sim 5 ~\mu m$, this corresponds to a total acoustic power of only $5~\mu W$, well within the cooling capacity of standard cryogenic setups. Recent advances in high-frequency SAW technology~\cite{SAW_45nm_Nelson_2012PRB, SAW_50G_Nelson, SAW_44G_2023}  allow operation in the tens-of-GHz regime, enabling us to explore wavelengths as short as $50$~nm. 
In addition, GHz SAWs on LiNbO$_3$ have been experimentally shown to possess coherence lengths in the centimeter range~\cite{SAW_long_coherence_length}, orders of magnitude larger than the few-micron device sizes considered here.
\begin{figure}[t]
  \centering
          \includegraphics[width=0.45\textwidth]{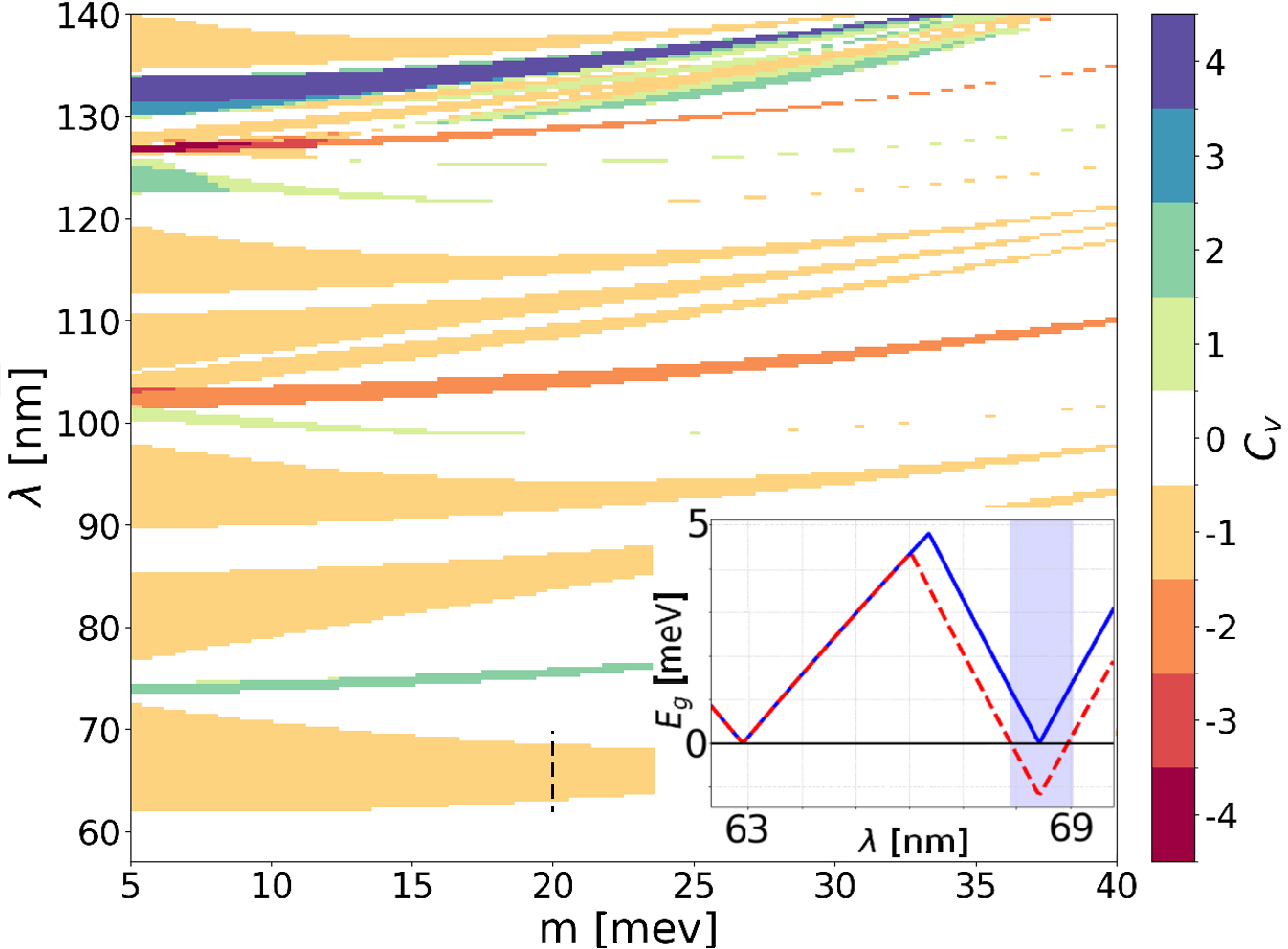}
  \caption{Support of $C_v$, the valley Chern number of the top valence miniband, in the $m$-$\lambda$ plane for $P=1$~W/m. Inset: direct and indirect gaps between the top valence miniband and bottom conduction miniband along the dotted line. A topological valley Hall response appears for $63~\textrm{nm} \lesssim \lambda \lesssim 68~\textrm{nm}$, where the indirect gap is positive.}
  \vspace{-6pt}
  \label{fig: Cv}
\end{figure}

The band structure is obtained by diagonalizing $H'_{\bm{k}}$ throughout the mBZ.
We find that SAWs can induce band inversions, as shown in Fig.~\ref{fig: band evolution}. 
At short wavelengths, the mBZ is large and the massive Dirac cone is only weakly perturbed. 
Increasing the SAW wavelength $\lambda$ strengthens the coupling, leading to sequential band inversions and topological phase transitions. 
A first gap closing at $\Gamma$ near $\lambda \approx 63$~nm gives the highest valence miniband a valley Chern number $C_v=-1$. A second closing at $M$ near $\lambda \approx 66$~nm returns it to a topologically trivial phase.

For massive graphene without SAWs, the continuum Dirac model predicts valley Chern numbers $C_v = \pm 1/2$ per spin. 
With a superlattice, however, the miniband periodicity requires integer $C_v$. 
As shown in the Berry-curvature panels beneath each spectrum, this is achieved through additional curvature that develops near the mBZ boundaries and modifies the original Dirac-point contribution, ensuring integer-valued $C_v$ for each miniband~\cite{Tan2024}. 
Across the two band inversions shown in Fig.~\ref{fig: band evolution}, the curvature remains highly localized in momentum space as seen in Fig.~\ref{fig: band evolution}J.
The momentum dependence of the Berry curvature is considered an important factor in determining the type of correlated state that emerges for nearly dispersionsless bands and in generalized Landau Level constructions~\cite{Yu2025, Liu2024GeneralizedLandauLevels}. Additionally, a large variance of the Berry curvature can lead to an enhanced Drude weight in metallic flat bands~\cite{Antebi2024,Mao2025}.
Finally, since $\Delta_{\bm{k}}$ only weakly breaks TR symmetry, the two valleys remain approximately related by TR, carrying equal and opposite valley Chern numbers and preserving their relative topology.

Fig.~\ref{fig: Cv} shows $C_v$ of the highest valence miniband as a function of $\lambda$ and $m$ for $P=1$~W/m. At short wavelengths $C_v=0$, but increasing $\lambda$ drives sequential band crossings where $C_v$ first reaches $-1$ and subsequently changes value multiple times as additional inversions occur. 
While we focus on the highest valence miniband, similar gap closings also appear in more remote bands.
A topological valley Hall response requires that the indirect gap remains positive across the mBZ; an example for $m=20$~meV is shown in the inset of Fig.~\ref{fig: Cv}. 
Finally, although occupations are set by lab-frame energies rather than quasienergies, this distinction is negligible for massive graphene.

The oscillatory behavior of the energy gap $E_g$ versus $\lambda$, shown in the inset of Fig.~\ref{fig: Cv}, appears for all values of $m$ and $P$.  
For small $\lambda$, up to the first gap closing, $E_g$ is well described by a minimal ten-band model that retains only the four nearest neighbors on the K-lattice for each momentum in the mBZ. 
As shown in~\cite{Note1}, this model predicts the first gap closing at $g_c=\sqrt{\lp v q \rp^2 + m^2-\Omega^2}$.

At the opposite long-wavelength regime, the mBZ shrinks and produces a dense set of \emph{atomic-like} minibands. One might expect that a superlattice varying slowly on atomic scales would simply fold the massive-graphene dispersion. Instead, the simultaneous growth of the piezoelectric coupling $g_{\mathrm{PE}} \propto \sqrt{P\lambda}$ and the shrinking of the mBZ generate well-separated and increasingly flat minibands.
Figure~\ref{fig: dense bands}A shows the bandwidth (BW) as a function of $\lambda$ for several values of $m$ and 
$P$. For fixed SAW power, increasing the mass suppresses the kinetic energy near the Dirac point and narrows the minibands; conversely, for fixed $m$, increasing $P$ enhances $g$ and further reduces the BW.
The resulting minibands can be extremely narrow, well below 1 meV before the first band inversion, as highlighted in Fig.~\ref{fig: dense bands}B and the atomic-like spectra in Fig.~\ref{fig: dense bands}C.
Beyond the first inversion the BW becomes non-monotonic, and nontrivial Chern numbers may also appear away from charge neutrality, as illustrated in Fig.~\ref{fig: dense bands}D.

Given the extreme flatness of these minibands, interaction effects are  enhanced. A conservative estimate of the Coulomb scale $E_c \approx 2$~meV (for a dielectric constant $\varepsilon_r = 4$),  gives an interaction-to-bandwidth ratio $E_c/BW \gtrsim 2$, comparable to values in magic-angle twisted bilayer graphene
\cite{MATBG_ee_Bultinck_2020, MATBG_ee_Xie_2020, MATBG_ee_Ledwith_2021, MATBG_ee_Song_2022, MATBG_ee_Xiao_2025}. We therefore anticipate that ASL flat bands may likewise host a range of correlated states.

\emph{Discussion---} We introduced a new class of periodic structures, the \emph{acoustoelectric superlattices}, and analyzed their distinctive electronic and topological properties. 
Our analysis has focused on a simple configuration in which the superlattice is generated by two orthogonal SAWs with identical wavelength and power. However, much richer superlattice architectures are possible. By varying the number of SAWs and tuning their power, wavelength, phase, and propagation direction, a large family of complex potential landscapes can be designed. 
This flexibility opens the door to engineering band structures tailored to support targeted electronic and topological phases, including highly correlated states.

Other 2D materials could likewise serve as electronic platforms for ASLs, broadening the range of accessible band-structure phenomena. 
Transition metal dichalcogenides are particularly appealing due to their large intrinsic band gaps and robust spin–valley locking.
Multilayer graphene offers additional opportunities, including tuning its spectrum via an external displacement field. An advantage of graphene-based systems is their exceptionally long mean free path, which makes them well suited for superlattices with large periodicities.

Finally, we expect ASLs will give rise to interesting topological transport responses. In particular, exploring whether ASLs with strongly broken TR symmetry can function as topological switches will be intriguing. 
These effects lie beyond the scope of the present work.

\begin{acknowledgments}
We thank Allan MacDonald, 
Johannes S.~Hofmann, 
and
Dominik M.~Juraschek
for illuminating discussions.  
TH acknowledges financial support by the 
European Research Council (ERC) under grant QuantumCUSP
(Grant Agreement No. 101077020). 
\end{acknowledgments}

\bibliography{references}

\end{document}


\title{
Supplementary Information for Miniband Generation by Surface Acoustic Waves
}
\author{Eli Meril}
\author{Unmesh Ghorai}
\author{Tobias Holder}
\author{Rafi Bistritzer}
\email{rafib@tauex.tau.ac.il}
\affiliation{School of Physics and Astronomy, Tel Aviv University, Tel Aviv 69978, Israel}
\date{\today}

\maketitle

\onecolumngrid

\section{Moving frame}
We seek a unitary transformation that maps the time-dependent Hamiltonian
$H_{\bm{k}}(t)$ in Eq.(2) of the main text to a time-independent form. For a general unitary operator $U_{\bm{k}}$, the Hamiltonian takes the form  
\begin{equation}
H'_{\bm{k}} = U_{\bm{k}} H_{\bm{k}} U_{\bm{k}}^\dagger +i \frac{\partial U_{\bm{k}}}{\partial t} U_{\bm{k}}^\dagger,          \label{Hkp}
\end{equation}
and the wavefunction transforms as $|\psi'_{\bm{k}}(t)\ra = U_{\bm{k}}(t)|\psi_{\bm{k}}(t)\ra$ ensuring that it satisfies the time-dependent Schrödinger equation in the new frame.
We consider an operator of the form 
\begin{equation}
U_{\bm{k}}(t) = e^{i F_{\bm{k}}(t)},
\qquad
F_{\bm{k}}(t) = \sum_{\bm{G}} \alpha_{\bm{k+G}}(t) \, 
c_{\bm{k+G}}^\dagger c_{\bm{k+G}}
\label{eq:Fk}
\end{equation}
where the sum runs over reciprocal lattice vectors of the superlattice. The time-dependent coefficients $\alpha_{\bm k+G}(t)$ are chosen to make $H'_{\bm{k}}$ time-independent. Using the Baker-Campbell-Hausdorff identity, we find 
\be
U_{\bm{k}} c_{\bm{k+G}} U_{\bm{k}}^\dagger =  c_{\bm{k+G}} \exp{\left[-i \alpha_{\bm{k+G}}(t)\right]}
\ee
which allows us to express $H_{\bm{k}}(t)$ explicitly in terms of the $\alpha$ coefficients. The phase factor of the hopping term associated with the first SAW is then 
\begin{equation}
\phi_1(t) = -\Omega_1 t - \alpha_{\bm{k+G}}(t) + \alpha_{\bm{k+G+q}_1}(t),      \label{phi1}
\end{equation}
with an analogous expression for the $\bm{q}_2$ term.
The time dependence is removed by requiring that  
\begin{equation}
\alpha_{\bm{k+G+q}_j}(t) - \alpha_{\bm{k+G}}(t)  = \Omega_j t, \qquad j=1,2  \label{alpha_eqs}
\end{equation}
Solving ~\eqref{alpha_eqs} gives  
\begin{equation}
\alpha_{\bm{k+G}}(t) = \left( i \Omega_1 + j \Omega_2 \right)t
\label{eq:alpha_sol}
\end{equation}
where the integers $i, j$ are given by the relation $\bm{G} = i \bm{q}_1 + j \bm{q}_2$.
The solution is independent of $\bm{k}$. 
The reference point for indexing $i,j$ is a gauge choice that only shifts the origin of the K-lattice without affecting physical observables.  
Substituting ~\eqref{eq:alpha_sol} into ~\eqref{eq:Fk} yields  $ U_{\bm{k}} = \exp{\lp i\Delta_{\bm{k}} t \rp}$ where $\Delta_{\bm{k}}$ is defined in Eq.(3) of the main text.

\section{SAW-induced potential}

We derive the scalar potential $V(x,t)$ induced in a two-dimensional material by a Rayleigh-type SAW propagating along the x-direction on a piezoelectric substrate located at $z=0$. 
The SAW couples to the 2D electrons through two mechanisms: a piezoelectric field and a deformation potential arising from strain transfer
\begin{equation}
V(x,t) = g \cos(qx - \Omega t), \quad g = g_{\mathrm{PE}} + g_{\mathrm{DP}},
\end{equation}
where $g_{\mathrm{PE}}$ and $g_{\mathrm{DP}}$ are the piezoelectric and deformation potential amplitudes, respectively. 
As discussed in the main text, the device geometry suppresses strain transfer, so that the coupling is dominated by $g_{\mathrm{PE}}$ and $g_{\mathrm{DP}}$ can be neglected.

The SAW generates an electric potential $\phi(x,z,t)$ that extends into both the substrate and the surrounding medium. Near the surface, we approximate it as 
\begin{equation}
\phi(x,z,t) = \phi_0 e^{i(qx - \Omega t)} e^{-q b(z) |z|},
\end{equation}
where $b(z)=1$ outside the substrate and $b(z)=b_s$ inside it.
While a full solution of the piezoelectric equations would self-consistently couple the displacement and the electric fields \cite{SAW_Zhang}, here we adopt a simplified approach. We treat the electromechanical coefficient $K_{\mathrm R}^2$ as an input parameter and fix $\phi_0$ by requiring that the electric‐field carried a known fraction $K_{\mathrm R}^2$ of the total SAW energy. 

The total SAW energy per unit area is related to the acoustic power per unit width $P$ and SAW velocity by $ U = P/v_s$. A fraction $K_{\mathrm R}^2$ of this energy is stored in the electric field,
\begin{equation}
U_E = K_R^2\,\frac{P}{v_s}\,.
\end{equation}
On the other hand, using $\bm E=-\nabla\phi$ and integrating the field energy density over z gives
\begin{equation}
U_E = \frac{1}{2} \int_{-\infty}^{\infty}\!\!dz\,\varepsilon(z)\,\lvert \bm E(z)\rvert^2 =
\frac12\,\varepsilon_{\rm eff}\,q\,\lvert\phi_0\rvert^2,
\end{equation}
where the effective permittivity is $\varepsilon_{\rm eff} = \frac{1}{2}(\varepsilon_{\rm env}+b_s\,\varepsilon_{\rm HF}) $, with $\varepsilon_{\rm env}$ the dielectric constant of the spacer and $\varepsilon_{\rm HF}$ the high-frequency permittivity of the substrate material. Equating the two expressions for $U_E$ and using $g_{\mathrm{PE}} = e|\phi_0|$ we obtain Eq.~(1) of the main text for $g_{\mathrm{PE}}$ and $\alpha_{\mathrm{PE}}$. 

We neglect screening by the 2D layer, as we focus on cases where the chemical potential lies within gaps or on strongly correlated flat bands that screen poorly.

\begin{figure}[h!]
    \centering
    \includegraphics[width=7cm]{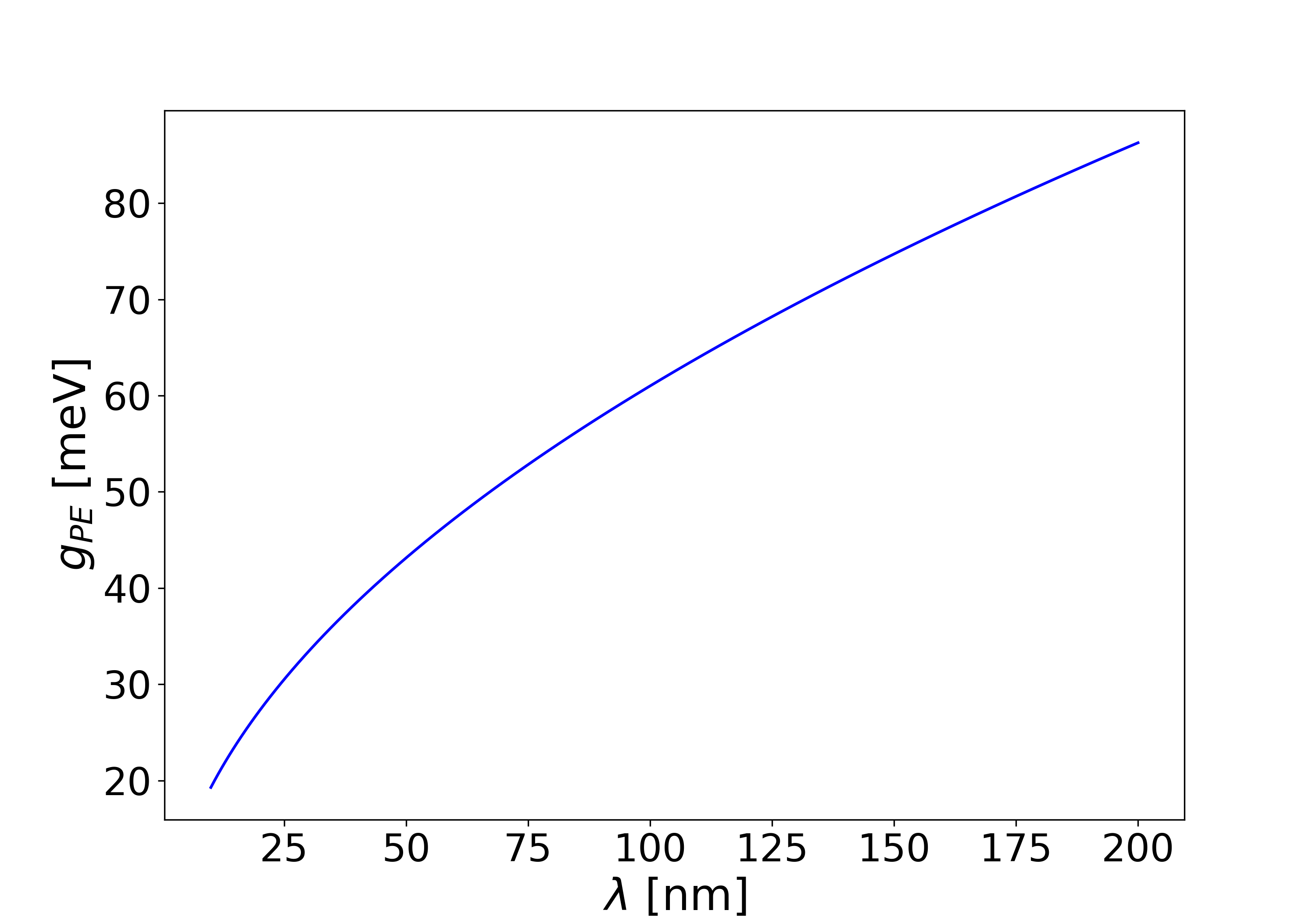}
    \caption{{\bf Piezoelectric coupling.} The coupling constant $g_{\mathrm{PE}}$ as a function of wavelength for $P = 1$ W/m, for a LiNbO$_3$ substrate and a hBN spacer.}
    \label{fig: inducedPotential}
\end{figure}

The coefficient $\alpha_{\mathrm{PE}}$ depends on the choice of materials, crystal cut, and device geometry. 
An optimal design maximizes piezoelectric coupling while minimizing strain transfer to the 2D material. For a representative device consisting of a  LiNbO$_3$ substrate and an hBN spacer, we use $v_s = 3488 \ \mathrm{m/s}, \quad \varepsilon_{\mathrm{HF}} = 40\varepsilon_0, \quad K_R^2 = 0.05$, and $b_s=0.6$,  with $\varepsilon_{\rm env}=4\varepsilon_0$. Expressing $g$ in meV, $P$ in W/m and $\lambda$ in nm, gives $ \alpha_{\mathrm{PE}} =  6.1 $. The resulting wavelength dependence of the coupling strength is shown in Fig.~\ref{fig: inducedPotential}.

\section{10-band model}

\begin{figure*}[htp]
    \centering
    \begin{minipage}[b]{0.5\linewidth}
        \centering
        \raisebox{0.5cm}{\includegraphics[width=0.5\linewidth]{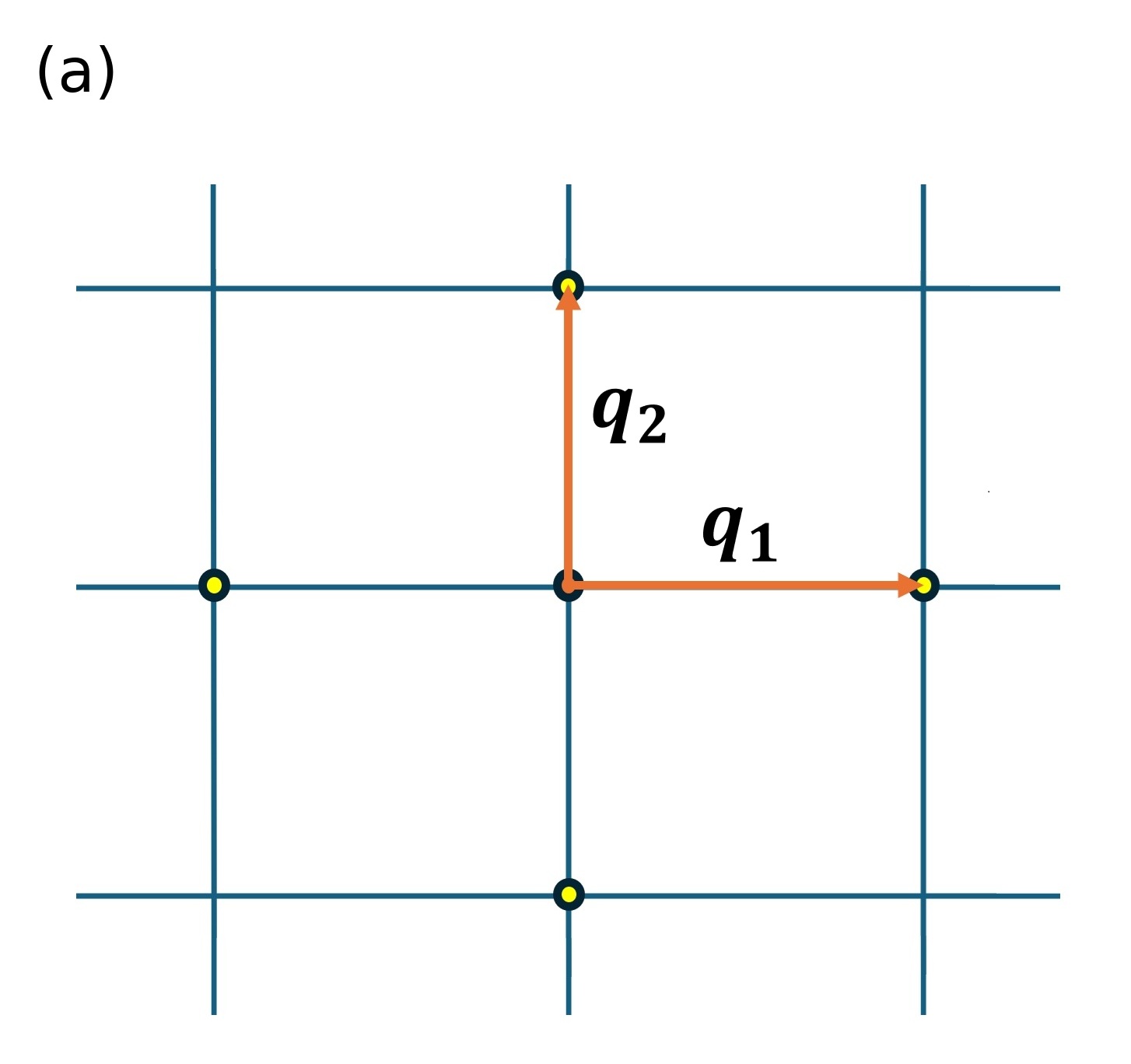}}
    \end{minipage}
    \hspace{-1.5cm} 
    \begin{minipage}[b]{0.5\linewidth}
        \centering
        \includegraphics[width=0.9\linewidth]{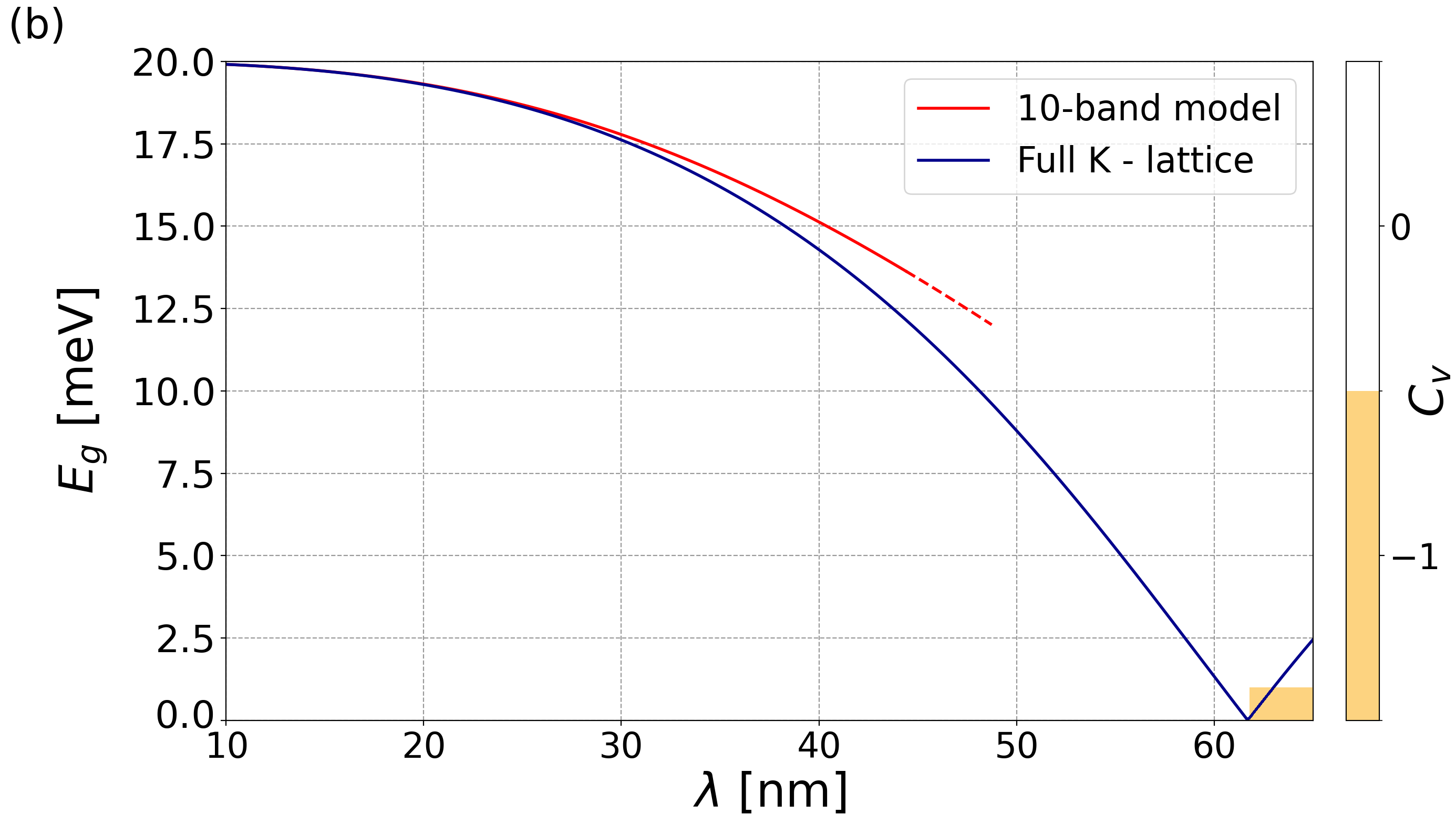}
    \end{minipage}
    \caption{(a) Truncated K-lattice used in the 10-band model, consisting of the central k-site and its four nearest neighbors. (b) Energy gap  $E_g(\lambda)$ between the valence and conduction minibands at the Dirac point, computed for $P = 1$~W/m and $m = 10$~meV, using the full K-lattice and the truncated 10-band model.}
    \label{fig: ten_band_schem}
\end{figure*}

In the limit of weak SAW coupling $g_j/vq \ll 1$, the problem simplifies considerably. In this regime, the wavefunction amplitude is primarily concentrated at momentum $\bm{k}$ in the mBZ and its four nearest neighbors on the K-lattice, which are connected by the SAW wavevectors $\pm \bm{q}_1$ and $\pm \bm{q}_2$ (see  Fig.~\ref{fig: ten_band_schem}a). Amplitudes on more distant K-lattice sites are negligible. Because each momentum point has two graphene sublattice degrees of freedom, this truncation yields a 10-band Hamiltonian. 

In this weak coupling regime, our numerical calculations show that the energy gap $E_g$ between the conduction and valence minibands first closes at the Dirac point.  Motivated by this observation, we use the 10-band model to analytically determine $g_c$, the minimum coupling strength at which this gap vanishes. We then benchmark the analytic result by comparing $E_g$ obtained from the truncated 10-band model with a full numerical calculation that includes the entire K-lattice.  

We consider two orthogonal SAWs of equal amplitude $g$ and equal wavevector magnitude $\bm{q}_1 = q \bm{\hat{x}}$ and $\bm{q}_2 = q \bm{\hat{y}}$. Using the conventions of the main text, the time dependence of the Hamiltonian is removed by transforming to the moving frame, yielding
\be
H'_{\bm{k}} = \sum_{\bm G} \left[ \lp h_{\bm{k+G}} - \omega_{\bm G}  \rp c_{\bm k+G}^\dagger c_{\bm k+G}  + \frac{g}{2} \lp c_{\bm k+G+q_1}^\dagger c_{\bm k+G} + c_{\bm{k+G+q_2}}^\dagger c_{\bm k+G} + \text{h.c.} \rp \right]   \label{full_ham_of}
\ee
where $\bm{k}$ is confined to the mBZ, $\Omega=v_s q$ is the SAW frequency, and  
$ h_k=v\,(\tau\sigma_xk_x+\sigma_yk_y) + m\sigma_z$. 
At the Dirac point and in the weak-coupling limit, \eqref{full_ham_of} reduces to a $10 \times 10$ Hamiltonian
\be
\cH_0=\begin{pmatrix}
        h_0&t&t&t&t\\
        t&h_{r} - \Omega &0&0&0\\
        t&0&h_{l} + \Omega&0&0\\
        t&0&0&h_{b} + \Omega &0\\
        t&0&0&0&h_{t} - \Omega
    \end{pmatrix}   \label{eig_eq}
\ee
where $h_0, h_r, h_l, h_b, h_t$ are the massive Dirac Hamiltonians evaluated at momenta $0, \pm\bm{q}_1, \pm\bm{q}_2$. The off-diagonal coupling is $t = g/2 ~ \sigma_0$ where $\sigma_0$ is the identity matrix in the sublattice space. The Hamiltonian $\cH_0$ acts on five two-component spinors $\Psi=(\phi_0, \phi_1, \phi_2, \phi_3, \phi_4)$. 

The condition for the gap closing corresponds to finding zero-energy solution of $\cH_0 \Psi = 0$. This leads to five coupled $2 \times 2$  equations.
The second equation $ t \phi_0 + (h_r-\Omega)\phi_1 = 0 $ can be solved to express $\phi_1$ in terms of $\phi_0$,
\be
\phi_1 = \frac{t}{\Omega^2-m^2-v^2q^2}(\Omega\sigma_0+\tau vq\sigma_x+m\sigma_z)\phi_0
\ee
Analogous relations express $\phi_2,\phi_3,\textrm{and}\,\phi_4$ in terms of $\phi_0$.
Substituting these expressions into the remaining equation $ h_0 \phi_0 + t (\phi_1 + \phi_2 + \phi_3 + \phi_4) = 0$ yields 
\be
\lp 1-\frac{g_c^2}{v^2q^2+m^2-\Omega^2} \rp\sigma_z\phi_0=0
\ee
which is satisfied when $g_c=\sqrt{v^2q^2+m^2-\Omega^2}$. Thus, $g_c$ represents the minimum SAW coupling at which the gap $E_g$ between the conduction and valence minibands at the Dirac point first closes.

Fig.~\ref{fig: ten_band_schem}b compares the energy gap $E_g(\lambda)$ predicted by the 10-band model with the exact result from the full K-lattice calculation. The agreement is excellent for wavelengths up to $\lambda \approx 30$~nm, validating the truncation in the weak-coupling regime. At larger wavelengths, which correspond to stronger effective interactions, additional K-lattice sites must be included to accurately capture the miniband structure.

\bibliography{references}